\documentclass[12pt,preprint]{aastex}

\newcommand{\co}{\mbox{$^{12}$CO}}
\newcommand{\coa}{\mbox{$^{13}$CO~}}
\newcommand{\cob}{\mbox{C$^{18}$O}}
\newcommand{\kms}{\mbox{km s$^{-1}$}}
\newcommand{\htwo}{\mbox{${\rm H}_2$~}}
\newcommand{\cc}{\mbox{cm$^{-3}$}}

\begin{document}
\title{Magnetically Aligned Velocity Anisotropy in the Taurus Molecular Cloud}

\author{Mark Heyer\altaffilmark{1}, Hao Gong\altaffilmark{1,2},
Eve Ostriker\altaffilmark{2},
Christopher Brunt\altaffilmark{1,3}}

\altaffiltext{1}{Department of Astronomy, University of Massachusetts, Amherst,
MA 01003-9305; heyer@astro.umass.edu}
\altaffiltext{2}{Department of Astronomy, University of Maryland, College Park, MD 20742-2421; hgong@astro.umd.edu,ostriker@astro.umd.edu}
\altaffiltext{3}{School of Physics, University of Exeter, Stocker Road,
EX4 4QL, United Kingdom; brunt@astro.ex.ac.uk}

\begin{abstract}
Velocity anisotropy induced by MHD turbulence is investigated 
using
computational simulations and molecular line observations of the 
Taurus molecular cloud. 
A new analysis method 
is presented
to evaluate the degree and angle of
velocity anisotropy using spectroscopic imaging data of interstellar
clouds.
The efficacy of this method is demonstrated on 
model observations derived from 
three dimensional velocity and
density fields from the set of numerical MHD simulations 
that span a range of magnetic field strengths.
The analysis is applied to \co\ J=1-0 imaging of a 
sub-field within the Taurus molecular cloud.  Velocity anisotropy is 
identified that is aligned within $\sim$10$^\circ$ of the mean local 
magnetic field direction derived from optical polarization measurements. 
Estimated values of the field strength based on velocity anisotropy 
are consistent with results from other methods. When combined with new column density measurements for Taurus, our magnetic field strength estimate
indicates that the envelope of the cloud is magnetically subcritical. 
These observations favor strong MHD turbulence within the low density, sub-critical, 
molecular gas substrate of the Taurus cloud. 
\end{abstract}

\section{Introduction}
Dense, interstellar molecular clouds offer a unique and valuable laboratory
to investigate magneto-turbulent phenomena.  These clouds are expected to be 
fully turbulent systems with 
a very large dynamic range between driving and dissipation scales.
The degree of magnetic coupling to the turbulent flows has important 
implications for 
the nature
of gas dynamics and star formation within molecular clouds. 
A strong, well coupled field can 
affect the star formation efficiency in a cloud by reducing 
the amount of material that is 
susceptible to gravitational collapse and star formation, 
and also affect the scale at which collapse occurs (Mouschovias 1976; 
Vazquez-Semadeni
etal 2005).
Magnetic fields also strongly affect the degree of gas density
compression in shocks.  
Such shock-generated 
density perturbations may provide the seeds of protostellar cores and 
protoclusters. 
Given the potential impact of the magnetic field on the gas dynamics
of molecular clouds, 
it is imperative to 
measure (or estimate) magnetic field strengths and to 
develop accurate descriptions of magnetohydrodynamic (MHD) 
turbulence 
under conditions applicable in star-forming clouds.

Goldreich and Sridhar (1995, hereafter GS95) developed a theory for
strong, incompressible, MHD turbulence that provides definitive
predictions of the spectrum and anisotropy of velocity fields.
Wave-wave interactions are expected to shear the Alfv\'en wave packet in
the plane perpendicular to the mean field.  Correspondingly, 
wave
energy is more efficiently redistributed to smaller scales in the
direction perpendicular to the field than 
through
the cascade parallel to the
field.
GS95 propose that 
a critical balance is achieved between non-linear interactions and
wave propagation,  such that the time scales to transfer
energy along the two directions are comparable, 
\begin{equation}
\lambda_\parallel/v_A \sim 
\lambda_\perp/v  
\end{equation}
where $\lambda_\parallel$ and
$\lambda_\perp$ are the 
wavelengths
parallel
and perpendicular to the mean field and $v$ is the mean velocity
fluctuation at the scale 
of the corresponding component. 
For 
an energy-conserving cascade,
$v \propto {\lambda_\perp}^{1/3}$, so equation (1) implies
\begin{equation}
\lambda_\parallel \propto {\lambda_\perp}^{2/3} 
\end{equation}
The
corresponding velocity scaling law along the magnetic field is
$v \propto {\lambda_\parallel}^{1/2}$.  
A critically balanced Alfv\'enic cascade
leads to a scale-dependent anisotropy of the velocity field.
This anisotropy has been demonstrated with
computational simulations for both incompressible and compressible MHD
turbulence (e.g. Maron \& Goldreich 2001; Cho, Lazarian, \& Vishniac 2002;
Vestuto, Ostriker, \& Stone 2003).

Can MHD induced velocity anisotropy, as predicted by GS95, 
be measured in interstellar clouds?  
Watson etal (2004) and Wiebe \& Watson (2007) have attributed the 
polarization properties of both OH masers and thermal molecular line 
emission 
to directionally dependent optical depths induced by MHD turbulence.
More panoramic observational views 
of the gas dynamics rely on spectroscopic imaging data of atomic or 
molecular line emission,
most notably, the HI 21cm line, and the low rotational transitions of 
\co\, and its isotopomers, \coa, and \cob.
In principle, the spatial variation of line shapes and velocity 
displacements offer a proxy view 
of the prevailing cloud dynamics.     
Recovering the form of the velocity power spectrum or its equivalent 
structure 
function from the spectroscopic data cubes, T(x,y,v), is 
challenging, owing to the complex integration of the velocity and 
density fields along the line of sight and the 
effects of line 
excitation and opacity that may filter or mask dynamical information from 
some fraction of the volume (Brunt \& Mac Low 2004; Ossenkopf etal 2006).  

Principal Component Analysis (PCA) is a 
powerful method to examine 
spectroscopic imaging data of interstellar clouds.  It 
reorders the data 
onto a set of eigenfunctions and eigenimages
(Heyer \& Schloerb 1997; Brunt \& Heyer 2002).  Characteristic 
velocity differences, ${\delta}v$, and spatial scales, $\tau$,
are derived for each
principal component for which the signal variance is distinguished 
from the statistical noise of the data.  The set of ${\delta}v,\tau$ 
points can be  empirically linked to the true velocity 
structure function 
using model velocity and density fields (Brunt \& Heyer 2002; Brunt etal 2003).
The method has been applied to 
a large set of \co\ and \coa\ imaging observations of giant molecular clouds
located within 4 kpc of the Sun to establish the universality of 
turbulence within the molecular interstellar medium (Brunt 2003; Heyer \& 
Brunt 2004). 
However, these studies did 
not consider velocity anisotropy.
The eigenvectors were derived from the 
covariance matrix that was accumulated from all spectra within the 
data cube with no orientation constraints.  Therefore, 
any dynamical signature 
of
anisotropy along a given axis 
was necessarily 
diluted by isotropic contributions to the covariance matrix.  The corresponding 
eigenimages identified locations within the projected plane 
where velocity differences can 
occur but at any angle. 

In this paper, we describe a modified application
of 
PCA
on spectroscopic imaging data to 
recover structure functions along perpendicular axes (\S2). 
In \S3, the utility of this analysis 
is demonstrated on spectroscopic data cubes derived from model velocity and 
density fields from decaying MHD simulations covering 
a range of magnetic field strengths.
In \S4, we apply this analysis to \co\ J=1-0 observations of 
a sub-field within the Taurus Molecular Cloud to show that such 
anisotropy is present and that the degree of anisotropy 
provides a coarse estimate to the strength of the 
magnetic field in this region.

\section{Description of Analysis Method: Axis Constrained Covariance Matrix}
To examine the degree of velocity anisotropy in interstellar clouds, we
have made a simple modification to 
the application of Principal Component Analysis.  
A directional constraint 
is imposed on the eigenvectors by calculating the covariance matrix from 
the sequence of spectra along one spatial axis 
(position-velocity slices of the data cube).  The position-velocity image 
can be extracted one slice at a time to preserve spatial resolution or 
can be generated 
by averaging contiguous slices to increase the signal to noise ratio. 
For a given data cube,
$T(x,y,v)$, with dimensions $n_x,n_y,n_v$, 
the position-velocity slice along the $x$ direction, averaged over thickness 
${\Delta}$
in the $y$ direction,
is
\begin{equation}
 W_y(x,v) = {{1}\over{\Delta}}\sum_{j=j1}^{j2} T(x,y_j,v) 
\end{equation}
where $\Delta=j2-j1+1$.
The covariance matrix for this position-velocity 
slice, 
${\bf C}^x$,
has components
 
\begin{equation}
C^x_{kl} = {{1}\over{n_x}}\sum_{i=1}^{n_x} W(x_i,v_k) W(x_i,v_l), 
\end{equation}
(suppressing the $y$ subscript on $W$).
The eigenvalue equation is solved for this axis-constrained covariance 
matrix,
\begin{equation}
 {\bf C}^x u_x = {\lambda_x}u _x  
\end{equation}
to produce the set of $n_v$ eigenvectors, $u_x(v)$, that describe 
velocity differences {\it exclusively} 
along this particular position-velocity 
slice.  To spatially isolate where these differences occur for each 
component, the spectra are projected onto the corresponding eigenvector,
\begin{equation}
 I_x(x_i) = \sum_{k=1}^{n_v} W(x_i,v_k) u_x(v_k) 
\end{equation}
The eigenprojection, 
$I_x(x)$, has dimensions $n_x \times 1$. 
The characteristic velocity difference, ${\delta}v_x$, and scale, $\tau_x$,
are determined from the scale length of the normalized autocorrelation 
functions of $u_x(v)$ and  $I_x(x)$ respectively
(Brunt \& Heyer 2002).  
Typically, only 4 to 5 $({\delta}v,\tau)$ pairs can be extracted from the 
axis constrained eigenvectors and projections for a given position-velocity  
slice owing to the limited spatial dynamic range.
These steps are repeated for all position-velocity 
slices ($j1=1,1+{\Delta},1+2{\Delta}, ...,n_y-{\Delta}$) 
in the data cube to produce a composite 
set of $({\delta}v_x,\tau_x)$ 
pairs derived from $n_y/\Delta$ sets of eigenvectors and eigenprojections. 
Similarly,
to examine structure along the $y$ axis, these steps are applied 
to position-velocity slices along the $y$ direction averaged over 
$x-$thickness ${\Delta}=i2-i1+1$, 
\begin{equation}
 W_x(y,v) = {{1}\over{\Delta}}\sum_{i=i1}^{i2} T(x_i,y,v) 
\end{equation}
with 
covariance matrix 
\begin{equation}
 C^y_{kl} = {{1}\over{n_y}}\sum_{j=1}^{n_y} W(y_j,v_k) W(y_j,v_l) 
\end{equation}
A corresponding 
set of 
$({\delta}v_y,\tau_y)$ pairs 
are derived from $n_x/\Delta$ sets of eigenvectors, $u_y(v)$, and 
eigenprojections, $I_y(y)$.   
For each axis, we consolidate the $\tau$ values into one pixel wide bins
and calculate the mean and standard deviation of ${\delta}v$ values 
for each bin.
Power laws are fit to 
each set 
to derive a relationship between the magnitude 
of velocity differences in the line profiles and scale over which these 
differences occur,
when constrained to each
axis,
\begin{mathletters}
\begin{equation}
 <{\delta}v_x> = v_{\circ,x} \tau_x^{\alpha_x} 
\end{equation}
\begin{equation}
 <{\delta}v_y> = v_{\circ,y} \tau_y^{\alpha_y}. 
\end{equation}
\end{mathletters}
The PCA scaling exponents, $\alpha_x,\alpha_y$,
are empirically linked to the scaling exponents 
of the first order velocity structure function 
\begin{mathletters}
\begin{equation}
\gamma = 1.69\alpha - 0.54 \;\;\;\alpha \leq 0.67  
\end{equation}
\begin{equation}
\gamma = 0.93\alpha - 0.03 \;\;\;\alpha > 0.67   
\end{equation}
\end{mathletters}
(Brunt etal 2003).  
The first-order structure functions, 
${\delta}v_x = v_{\circ,x} \tau_x^{\gamma_x} $ and 
${\delta}v_y = v_{\circ,y} \tau_y^{\gamma_y} $ 
provide equivalent 
information to the power spectrum of the velocity field along the 
$k_x$ and $k_y$ axes, respectively, for $k_z=0$. 
Averaging over $\Delta$ in $y$ (or in $x$) is equivalent to
integrating along $k_y$ (or $k_x$).

This method offers a tool to derive velocity structure functions along 
any two perpendicular 
(spatial) 
axes of a spectroscopic data cube. 
With apriori knowledge of the local magnetic 
field direction, one could simply rotate the data cube to align the 
$x$-axis along this direction and determine the parallel and perpendicular 
structure functions.  
However, this orientation may not 
necessarily correspond to the angle at which velocity anisotropy is 
largest.  A more rigorous test of MHD-induced 
anisotropy is the demonstration that velocity anisotropy is {\it maximized}
when one of the two orthogonal axes lies
along the local magnetic field direction. 
To determine the angle of maximum anisotropy, $\theta_{MAX}$, the 
spectroscopic data cube is rotated through a sequence of angles, $\theta$,
in the plane of the sky from which the $x$ and $y$-axis structure functions 
are calculated for each angle.
To compare with polarization observations that measure 
position angles east of north, we define $\theta$ as the angle measured 
counter-clockwise from the $y$ axis.
To quantify the difference between the $x$ and $y$-axis 
structure functions for each angle, we consider two separate 
measures of anisotropy.   The first anisotropy index, $\Psi_1$, is 
motivated by GS95 who predict differences in the scaling exponents,
$\gamma_x,\gamma_y$,
\begin{equation}
\Psi_1 = {{\gamma_x-\gamma_y} \over 
         {\gamma_x+\gamma_y}} 
\end{equation}
Based on the results shown in 
\S3.1 and \S3.3, the 
second anisotropy index, $\Psi_2$, measures the difference 
between the normalization constants, $v_{\circ,x}, v_{\circ,y}$
\begin{equation}
\Psi_2 =  {{v_{\circ,y}-v_{\circ,x}} \over 
         {v_{\circ,y}+v_{\circ,x}}} 
\end{equation}
For an isotropic velocity field, $\Psi_1 \approx 0$ and 
$\Psi_2 \approx 0$. 

The modulation of $\Psi$ by position angle enables 
a more accurate determination of the amplitude and angle at which the velocity 
anisotropy is maximized.  This modulation involves rotation 
about an axis so there is degeneracy for angles $\theta$ and
$\theta+180$.  
We find that
the function 
\begin{equation}
\Psi (\theta) = \Psi_{\circ} cos [2(\theta - \theta_{MAX})]  
\end{equation}
provides a reasonable 
fit 
to the variation of the anisotropy index. 
The coefficient $\Psi_{\circ}$ gives the amplitude of the anisotropy 
and the phase, $\theta_{MAX}$, is the angle of maximum anisotropy that can 
be compared to the local field direction, $\langle \theta_B \rangle$. 

\section{MHD Simulations}
To demonstrate that the analysis described in \S2 can indeed recover the 
spatial statistics of the velocity field,
we have
analyzed a set of computational simulations of decaying MHD turbulence
from 
Ostriker, Stone, \& Gammie (2001).
The models span a range of magnetic field
strengths parameterized by the ratio of thermal to 
mean-field 
magnetic energy densities,
$\beta = c_s^2/v_A^2$ where $c_s$ is the sound speed for \htwo\ and $v_A$ is the
Alfv\'en velocity based on the mean field, 
$\langle {\bf B}\rangle /\sqrt{4\pi\rho}$.  In these decaying-turbulence
simulations, the initial velocity field is identical for all 
models, so any subsequent differences in the velocity 
field (including anisotropy) arise due to magnetic effects.
For each $\beta$ model
($\beta=0.01,0.1,$ and 1),
we examine two 
snapshots 
at times $t$ 
in units of the sound crossing time, $t_s$.
The snapshots are chosen such that the kinetic energy, or sonic 
Mach number ${\cal M}_s= v_{rms}/c_s$, is comparable for all $\beta$.
A summary of the simulation
snapshots
is listed in Table~1. 
The cloud models are initially threaded with a spatially uniform magnetic field along 
axis 1 of the volume (${\bf B}_\circ=(B_1,0,0)$);
owing to periodic boundary conditions adopted for the simulations, the
mean field $\langle {\bf B}\rangle={\bf B}_\circ$ at all times,
although the total magnetic field strength $|{\bf B}|$ changes in time.
Depending on the strength of the mean magnetic 
field, $B_\circ$, turbulent flows can distort the magnetic field lines.
A coarse estimate to the large scale magnetic field alignment is provided by 
the variance of each component $B_1$, $B_2$ and $B_3$ normalized by 
the total field 
$B = (B_1^2 +B_2^2 + B_3^2)^{1/2}$.  For the strong field simulation 
($\beta=0.01$; snapshots B2,B3),
the variance of the 
magnetic field components on each axis is small ($<$ 3\%), 
indicative of a spatially 
rigid field well 
aligned along axis 1 for most cells within the volume.
For the intermediate ($\beta=0.1$; snapshots C2, C3) and weak field 
($\beta=1$; snapshots D2,D3) cases, the 
fluctuations of the field values are significantly larger (10-40\%),
reflecting localized tangling 
of the magnetic field.  

\subsection{Direct Measurement of Velocity Anisotropy}

The MHD velocity anisotropy is manifest by the spectral properties of the 
velocity field along axes 
parallel and 
perpendicular to the 
magnetic field.  Cho, Lazarian, \& Vishniac (2002) and Vestuto, Ostriker, \& 
Stone (2003) examined the spectral slopes of directional power spectra
or equivalently, the 2nd order structure function, of velocity fields
from computational simulations.  
Both studies found steeper spectral 
slopes and smaller normalization constants for structure functions 
extracted along the magnetic field direction relative to those along 
an axis perpendicular to the field.  
That is, the velocity field contains more power when $k_\perp\approx k$
and $k_\parallel \approx 0$ than
when $k_\parallel \approx k$ and $k_\perp \approx 0$, for a given $k$.

\begin{figure}[ht]
\begin{center}
\includegraphics[width=0.95\hsize]{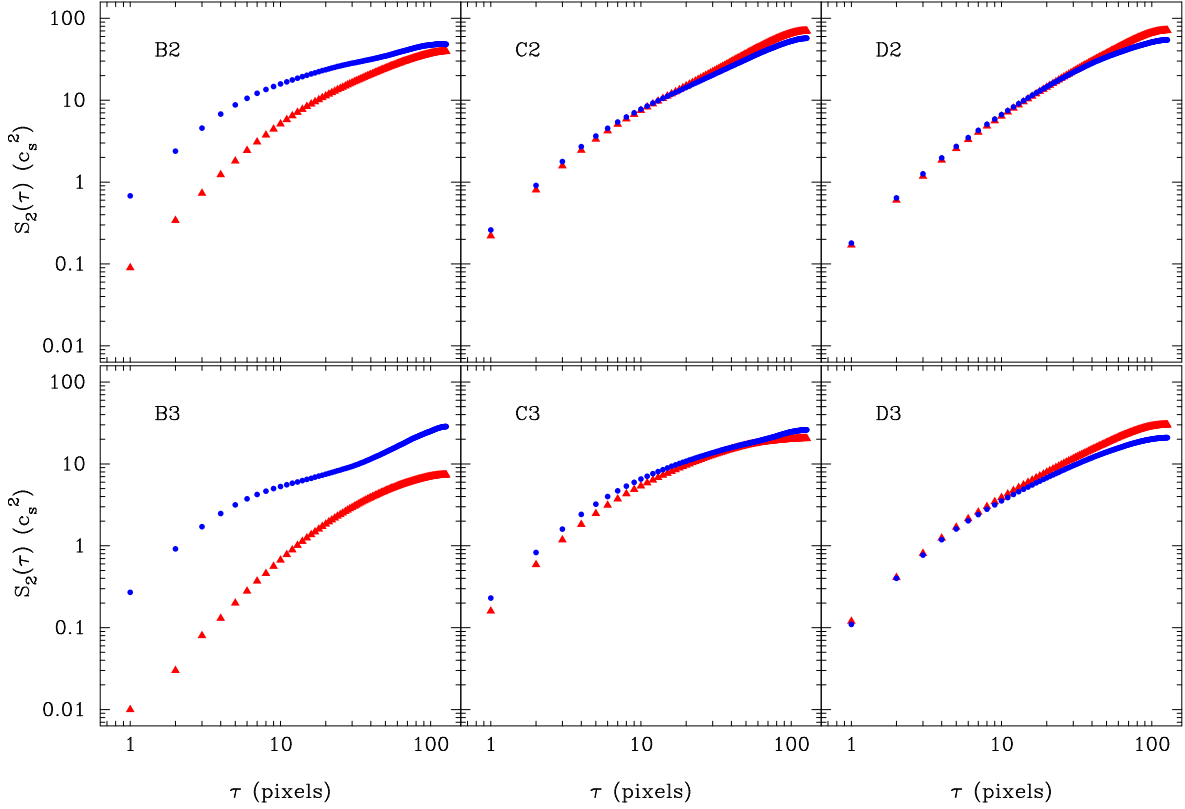}
\caption{The second order velocity structure function, $S_2(\tau)$,
along axes parallel (red triangles) and perpendicular (blue circles) to the 
mean 
magnetic field direction for 
velocity fields 
from turbulent simulations.
Velocity anisotropy is 
evident in the strong field model
cases (B2, B3) as a larger scaling amplitude and shallower index for 
the structure function perpendicular to the mean field direction.}
\label{fig1}
\end{center}
\end{figure}

To quantify the 
velocity anisotropy in the simulations used in this study 
and to compare with simulated observations
shown in \S3.2, 
the true 2nd order structure function, $S_2(\tau)$,
is calculated directly from 
each model velocity 
field, $v_2$, 
 
\begin{equation}
S_2(\tau_\parallel,\tau_\perp) 
= \langle [v_2(\mathbf{x})-v_2(\mathbf{x}+\mathbf{\tau})]^2\rangle
\end{equation}
  
where $\tau = \tau_\parallel\mathbf{e_\parallel}+\tau_\perp\mathbf{e_\perp}$;
$\mathbf{e_\parallel},\mathbf{e_\perp}$ are unit vectors parallel and 
perpendicular respectively to the local mean magnetic field direction, 
and the angle brackets 
denote a spatial average over the volume (Cho, Lazarian, \& Vishniac 2002). 
Here, we restrict our analysis to the projected plane appropriate for the 
velocity field to facilitate comparison with model observations in \S3.2. 
In this case, the $v_2$ component projects into the plane
\footnote{
To distinguish between the 3 spatial coordinate axes of the models and the 
observed projected axes, we reference the 3 spatial axes of the model 
fields as 1,2,3 and 
label the projected, observed axes as $x$ and $y$.}
defined by axes 1 and 3.
Figure~\ref{fig1} shows the 2nd order structure functions, $S_2(0,\tau_\perp)$
and $S_2(\tau_\parallel,0)$. 
Power laws are fit over the pixel range 5-15 to exclude the steep 
component at small scales that results from 
grid-scale numerical dissipation
of the 
simulation.  The amplitudes and spectral indices of an equivalent, 
first order structure function,
$(S_2)^{1/2}$, are listed in Table~2. 
Velocity anisotropy is clearly identified in the 
B2 and B3 simulation snapshots as the slope and amplitude of the 
orthogonal structure 
functions are different. 
For the intermediate (snapshots C2,C3) and weak (snapshots D2,D3)
B-field cases, the structure functions are statistically equivalent, 
indicative of globally isotropic velocity fields with 
slopes ($\sim$0.5) that are typical 
of strongly supersonic, super-Alfvenic turbulent flows. 
The absence of velocity anisotropy results from 
the local distortions of the magnetic field that dilute any signature to large scale 
anisotropy.  

\subsection{Model Spectroscopic Data Cubes}
Observers do not directly recover the 
3 dimensional velocity fields.  Wide field
spectroscopic imaging measures line intensity as a function of position 
on the sky and velocity along an axis. The precise shape of a
line profile  is 
dependent on density, the projected velocity component, temperature, 
and chemical abundance that are  
integrated 
along the line of sight and affected by line excitation and opacity.
To place the model velocity and density fields from the computational 
simulations in 
the same domain as 
observations, we generate synthetic line profiles of 
\co\ and \coa\ J=1-0 emission.  
Details of the line excitation and radiative 
transfer calculations are described by 
Brunt \& Heyer (2002).  The assumed 
abundance values of \co\ and \coa\ relative to 
\htwo\ are 
1.0$\times$10$^{-4}$ and 1.25$\times$10$^{-6}$ respectively. 
We adopt a uniform kinetic 
temperature of 15 K, which corresponds to a one dimensional sound speed of 
0.22 \kms. 
The adopted mean volume density of \htwo\ is $n=1000$ \cc.

The choice of constructing 
synthetic profiles of the high opacity \co\  emission is motivated 
by two factors.  First, the \co\ J=1-0 line is the most common tracer 
of cloud structure so there are many observational data sets available 
to compare with these
models.  
To be sure, \co\ does not effectively probe the high density 
cores of molecular clouds where 
star formation takes place.  However, these regions comprise a small fraction of the cloud mass 
and volume (Heyer, Ladd, \& Carpenter 1996; Goldsmith etal 2008). 
Brunt \& Heyer (2002) examined the effects of 
line opacity on the gas dynamics perceived by observations.  With the exception of 
micro-turbulent velocity fields, they found that \co\ measurements 
reliably recover the velocity field statistics.  Although the local 
optical depth can be large within a volume,  
the macro-turbulent velocity fields provide an 
effective large velocity gradient condition that allows most 
photons from the surface of the local volume to escape.  In addition, 
owing to radiative trapping, \co\ is detected over a broader 
area than the lower opacity lines so 
there are simply more measurements and information on the largest scales.
Nevertheless, to re-examine the effects of line opacity, 
we also generate 
and analyze synthetic profiles of the \coa\ J=1-0 transition. 

\subsection{Axis-Constrained PCA Applied to Model Data Cubes}
The utility of the analysis described in $\S$2 is assessed by its 
application to the synthetic \co\  and \coa\ 
data cubes constructed from the MHD model density and velocity 
fields.  Does the analysis recover velocity anisotropy when this is 
present in the raw velocity field,
for the case 
of strong magnetic fields?  
Does the method verify isotropic 
velocity fields in the intermediate and weak field cases? 

\begin{figure}[ht]
\begin{center}
\includegraphics[width=0.95\hsize]{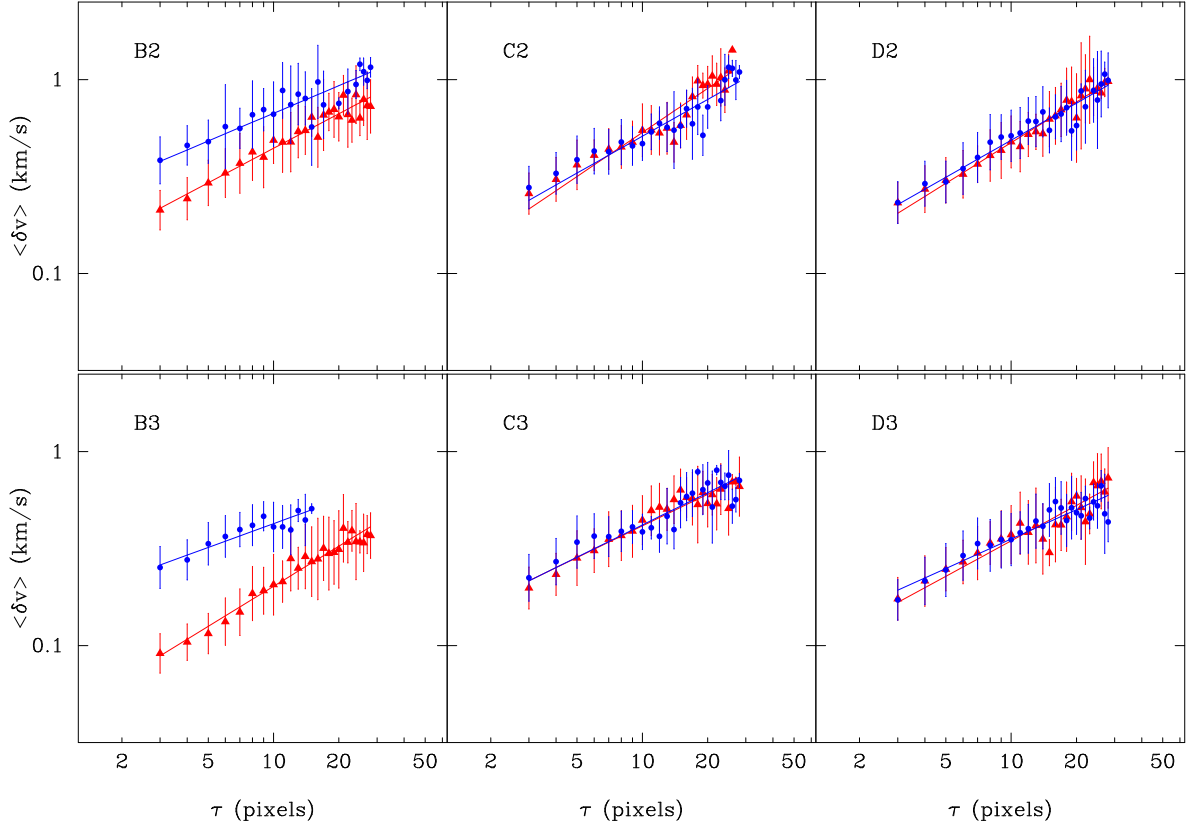}
\caption{The axis-constrained PCA 
$<\delta v>,\tau$ 
relationships derived from 
synthetic 
spectroscopic data cubes of \co\ J=1-0 
emission 
for the $x$ axis (
along the mean magnetic field;
red triangles) and $y$ axis 
(
perpendicular to the mean magnetic field;
blue circles).  
The error bars reflect the standard deviation of values 
within each 1 pixel wide bin of $\tau$.
The method recovers the 
anisotropy intrinsic to the B2 and B3 model 
velocity fields and verifies the isotropic velocity fields of 
the intermediate (C2,C3) and weak field (D2,D3) models.}
\label{fig2}
\end{center}
\end{figure}

Here, we examine the 
synthetic
spectroscopic data cubes derived from the 
$v_2$ velocity field 
for position angle $\theta$=90$^\circ$, i.e. corresponding to alignment of 
the $x$-axis with the mean magnetic field direction
in the plane of the sky.
The results of the axis-constrained PCA method (with 
$\Delta$=2), as applied to all model snapshots, 
are shown in Figure~\ref{fig2}.
Magnetically aligned anisotropy 
is clearly identified for the B2 and B3 simulation
snapshots
 as a separation 
of the set of points
$(<{\delta}v>,\tau)$ 
derived respectively along the $x$ and $y$ axes 
of the model data cubes.  This separation of points is qualitatively 
similar to the corresponding true structure functions calculated directly 
from the velocity fields that are shown in Figure~\ref{fig1}. 
For the intermediate- (C2,C3) and weak- (D2,D3) magnetic field 
snapshots,
there is a 
strong overlap of points $(<{\delta}v>,\tau)$ 
derived for the orthogonal axes.
This indicates velocity
isotropy with respect 
to the mean magnetic field, and is in agreement with the true velocity 
structure functions for the models. 

To assess the method quantitatively,
bisector fits of power laws with parameters, $\alpha$, $v_\circ$,
are fit to each set of points for each axis 
over the  range 
$3 \le \tau \le 30$ pixels. 
The scaling exponents, $\gamma_\parallel$ and $\gamma_\perp$, of the structure function are 
derived from the fitted parameters, $\alpha_\parallel$ and $\alpha_\perp$, according to 
equation 10.  The results for the \co\ and \coa\ model data cubes 
are summarized in Table~3.  
With the exception of the C3 model data cube, there 
are no significant differences between the power law parameters 
derived from \co\ and \coa\ model cubes,
demonstrating that opacity effects do not significantly skew the 
derived velocity field statistics.

For the strong field simulations, 
the separation of points in Figure~\ref{fig2} is due to 
a combination 
of a larger normalization constant and shallower index 
for the perpendicular structure function.  
Moreover, the anisotropy 
is stronger in the later stage simulation 
(comparing B3 with B2).
There are,
however,
discrepancies between the values of $\gamma$ determined directly from 
the velocity field in Table~2 and those determined by PCA that are listed in 
Table~3.  The root-mean-square difference between power law indices is 0.12 
(23\%). 
This discrepancy is due in part, to the difficulty in 
measuring a power law index of structure functions of 
velocity fields produced by the computational simulations that have limited 
inertial 
range (Vestuto, Ostriker, \& Stone 2003). 
In addition, the PCA eigenprojection along a single axis tends to limit the 
dynamic range of spatial scales over which the power laws parameters
are derived.    
Despite this discrepancy of the scaling exponents, these results
demonstrate the ability of the axis constrained PCA eigenfunctions to 
show a clear signature of 
velocity anisotropy induced by MHD turbulence.  

\section{The Taurus Molecular Cloud}
\begin{figure}[ht]
\begin{center}
\includegraphics[width=0.95\hsize]{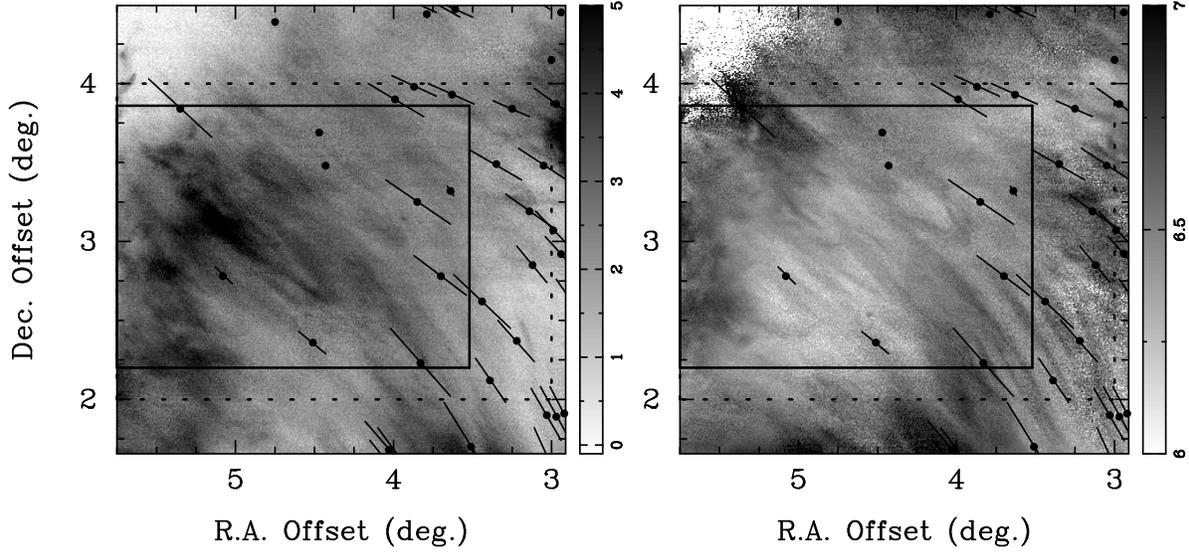}
\caption{(left) Image of \co\ J=1-0 emission of a sub-field within the 
Taurus molecular cloud integrated over the velocity interval 5.5 to 7.5 \kms\ 
and (right) image of \co\ velocity centroid 
(Narayanan etal 2008),
with overlay of 
optical 
polarization vectors from the compilation by Heiles (2000).  
The molecular line emission and velocities exhibit streaks that are aligned 
along the local magnetic field direction. 
The solid line box outlines the area upon which the
axis constrained PCA method is applied.  The dotted line 
box shows the area within which the polarization angles are 
averaged to esimate the mean magnetic field direction. 
}
\label{fig3}
\end{center}
\end{figure}

The Taurus Molecular Cloud provides a valuable platform to investigate 
interstellar gas dynamics and the star formation process, owing to its
proximity 
(140 pc)
and the wealth of complementary data.  
Narayanan etal (2008) present new wide-field imaging 
observations of \co\ and \coa\ J=1-0 emission from the central 100 deg$^2$
of the Taurus cloud complex, obtained with the FCRAO 14m telescope.  The 
images identify a low column density substrate of gas that contain 
subtle streaks of elevated \co\ emission
aligned along the local magnetic field direction as 
determined from stellar polarization
measurements (Heiles 2000).  
Images of \co\ J=1-0 integrated intensity and centroid velocity with
measured polarization vectors 
from this subfield 
are shown in Figure~\ref{fig3}.   These show a connection between the 
density and velocity fields. 
While the origin of these streaks is
unknown, their rigorous alignment with the polarization vectors 
strongly 
suggests that the interstellar magnetic field plays a prominent role
in the gas dynamics of this low density material. 

\begin{figure}[ht]
\begin{center}
\includegraphics[width=0.95\hsize]{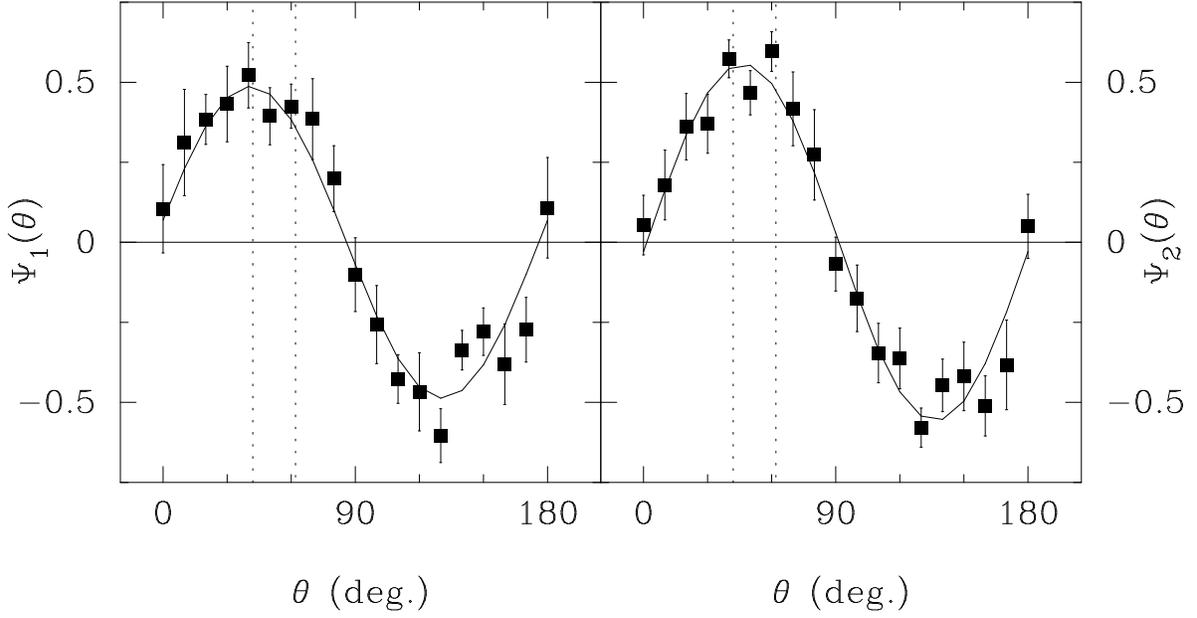}
\caption{The variation of the anisotropy indices,
$\Psi_1$ (left) and $\Psi_2$ (right), with 
position angle, $\theta$.  For each index, the solid line shows the fit of 
equation 13 to the set of points.
The dashed vertical lines show 
$\langle \theta_B\rangle \pm 1\sigma$ inferred from optical 
polarization measurements of background stars within the subfield.
The angle of maximum 
anisotropy is nearly aligned with the local magnetic field direction, which 
suggests a relationship between velocity 
anisotropy and the interstellar magnetic field induced by 
strong MHD turbulence.
}
\label{fig4}
\end{center}
\end{figure}

\begin{figure}[ht]
\begin{center}
\includegraphics[width=0.95\hsize]{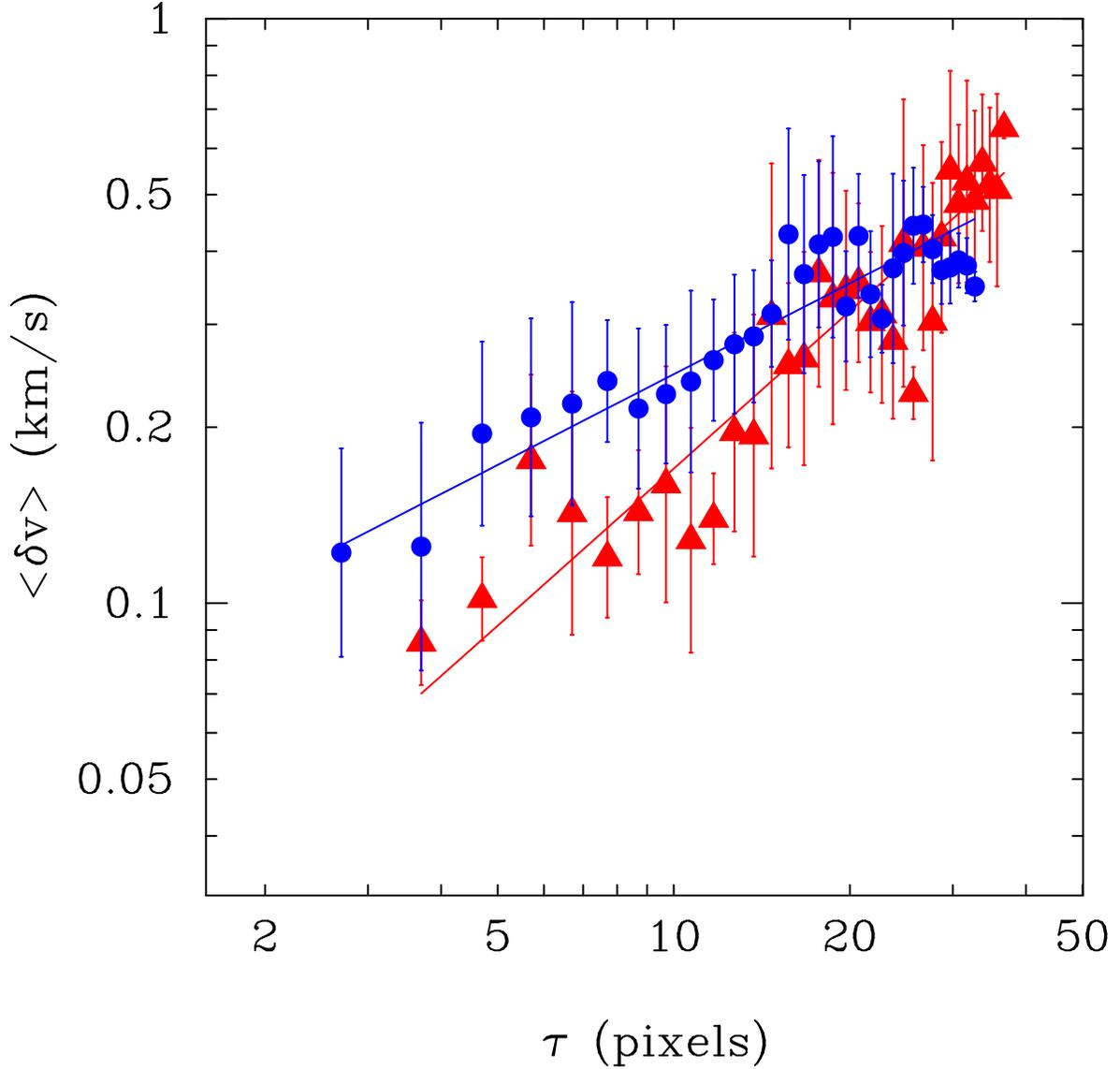}
\caption{The PCA derived 
$\delta v, \tau$ 
relationship derived from the 
\co\ 
spectroscopic data cube of the Taurus sub-field rotated to align the 
x-axis with the angle of maximum anisotropy, $\theta_{MAX}=46^\circ$.
The red triangles are points derived along the rotated x-axis and the 
blue circles are points derived along the rotated y-axis.  The error 
bars reflect the standard deviation of values in each bin. 
The pattern is similar to that found 
from the B2, B3 simulation snapshots
in Figure~\ref{fig2} and suggests an important role of the magnetic 
field on the local gas dynamics.
}
\label{fig5}
\end{center}
\end{figure}

To assess the degree of velocity anisotropy within this sub-region
of the Taurus molecular cloud, 
we have applied the axis constrained PCA method 
to
the \co\ data from this 
imaging survey.  The precise field is described by the solid box in Figure~3. 
We do not consider the \coa\ J=1-0 data since the signal is 
weak from this low column density sector of the cloud.  
The mean, local polarization angle,
derived from 16 measurements within the 
field 
is 52$^\circ\pm$10$^\circ$.  Assuming the polarization is induced
by selective absorption of background starlight by magnetically aligned,
elongated 
dust grains, this angle corresponds to the local magnetic field direction
(Purcell 1979; Draine 2003).
Figure~\ref{fig4} shows the variation of the anisotropy 
indices, $\Psi_1$ and $\Psi_2$, with position angle 
(measured east of north) for \co\ 
data within this subfield of the Taurus cloud.  
For $\Psi_1$, which considers the differences in scaling exponents,
the fitted parameters are 
$\Psi_{\circ}$=0.49$\pm$0.03 and
$\theta_{MAX}$=41$^\circ\pm$2$^\circ$.  
For $\Psi_2$, which measures anisotropy based on the differences 
of the normalization constants, 
$\Psi_{\circ}$=0.56$\pm$0.03 and
$\theta_{MAX}$=46$^\circ\pm$2$^\circ$.  
The angle of maximum anisotropy is
within 6-11$^\circ$ 
of the local magnetic field direction and the mean position angle of the 
emission streaks of \co\ emission. 
The $x$ and $y$-axis structure 
functions derived at $\theta_{MAX}$=46$^\circ$ 
are shown in Figure~\ref{fig5}. 
These 
distributions
show the same pattern of 
offsets between
the 
parallel and perpendicular structure functions measured
in the strong field simulation snapshots (B2,B3) shown in Figure~\ref{fig2}. 
For the Taurus field,
the power law index of the structure function derived from \co\ 
along the $x$-axis 
(i.e. the direction aligned with the polarization)
is 
steeper (0.81$\pm$0.05) than the index of the $y$-axis structure 
function (0.34$\pm$0.06).  
The steeper power law 
along the x-axis
is indicative of a 
velocity field more dominated by large scales.
Similar to the 
model structure functions 
in the strong magnetic field cases,
the normalization of the $y$-axis structure function, $v_{\circ,y}$ is
0.08 \kms\ and larger than the 
value of the $x$-axis structure function ($v_{\circ,x}$=0.02 \kms).
Thus, the smooth variation of density along the presumed magnetic
field is mirrored by a smooth variation in the velocity, and the
stronger variation in density in the perpendicular direction
(streakiness) is mirrored by a stronger variation in the velocity.
Indeed, preliminary analysis shows that in the direction perpendicular to the 
projected magnetic field, displacements between the peaks in 
integrated intensity and velocity centroids are similar with typical 
values 0.2 to 0.4 pc.

The results shown in Figures~3,4,5 are 
suggestive of velocity anisotropy induced by strong MHD turbulence,
as described by GS95 and verified by computational simulations
(Cho, Lazarian, \& Vishniac 2002; Vestuto, Ostriker, \&  Stone 2003). 
We note that the observed spectral slope parallel to the 
field, $\gamma_\parallel$,
is 
steeper than the value predicted for incompressible MHD turbulence by GS95 but 
is similar to values derived for the strong field (B2, B3) simulations.  
Velocity anisotropy 
could be produced by processes other than 
MHD turbulence.  A systematic flow of material 
that is ``channeled'' by the magnetic field may also generate 
differences in the parallel and perpendicular structure functions. 
Such large scale gradients would produce steep 
spectral indices ($\gamma \ge 1$).
However, the observed high frequency variation of velocities
perpendicular to the
field are not characteristic of such large scale shear flows.
Regardless of its origin, the near alignment of the velocity 
anisotropy with the local magnetic field direction demonstrates 
the importance of the interstellar magnetic field on the gas 
dynamics within this low density component of the Taurus molecular 
cloud.  

\subsection{The Magnetic Field Strength in the Taurus Cloud Envelope}
Since anisotropy is only 
evident in models with strong magnetic fields, the 
identification of
such anisotropy within observational data 
offers a proxy measure of the magnetic field and its 
effect 
upon the neutral gas 
(Vesuto, Ostriker, \& Stone 2003).   Specifically, the amplitude of
the mean magnetic field, 
$B_\circ=|\langle {\bf B} \rangle|=c_s \sqrt{ 4{\pi}{\rho}/\beta}$,
may be 
estimated from values of $\beta$ that are constrained by the
observations.
The measured degree of velocity anisotropy
is sensitive to the projected component of the mean
field in the plane of the sky.  It is improbable that
the magnetic field threading an interstellar cloud
is aligned in the sky plane.  Therefore, measures of velocity
anisotropy provide a lower
limit to the value of $B_\circ$.

Based on our analyses,
the anisotropy measured 
in the Taurus subfield is not as large as in the
strong field 
snapshots (B2,B3), but it is larger than the anisotropy limits 
for the intermediate-field
strength model snapshots (C2,C3).  
Given this 
bracketing, we can assign an approximate value of $\beta$=0.03 
to the Taurus subfield 
as a logarithmic midpoint between the intermediate 
and strong field models.  
Since the observed region is within the 
low column density regime of the Taurus Cloud, we set the kinetic 
temperature to be 15 K and the mean density to be 250 \cc.
These values for the temperature and density are reasonably constrained 
by non-LTE excitation models that match 
the observed \co\ and \coa\ J=1-0 intensities from 
the sub-thermally excited component of the Taurus cloud (Goldsmith etal 2008).
The magnetic field strength corresponding to these values of 
$\beta$, kinetic temperature, and gas density is 14 $\mu$G.  
As noted above, this is a lower limit on the total magnetic field
strength since the velocity anisotropy is not sensitive to the
line-of-sight component of the magnetic field.

Zeeman measurements of the OH 
line emission from the L1544 dark cloud, located $\sim$4 degrees to the south-west of the subfield in Taurus,
identify a line 
of sight field strength of 11 $\mu$G (Crutcher \& Troland 2000).  
While this value is comparable to our coarse estimate of the field,
these OH Zeeman observations are toward higher column 
density material ($N(H_2) \sim 2\times10^{22}$) than is likely present in the 
Taurus subfield.   If this higher column density reflects a larger 
volume density
and if the magnetic field is correspondingly compressed,
the field in the diffuse parts of the Taurus cloud may be smaller. 

The Chandrasekhar \& Fermi (1953) method offers an additional measure of the 
magnetic field strength in interstellar clouds.  It attributes 
deviations of the local magnetic field from the mean field direction
to linear-amplitude transverse MHD waves such that 
\begin{equation}
 ({\delta}B/B_p) = |{\delta}v|/v_A  
\end{equation}
where $B_p$ is 
the projection of the mean magnetic field on 
the plane of the sky, ${\delta}B$ and $|{\delta}v|$ are components of 
the magnetic and 
velocity perturbations transverse to $B_p$, and $v_A$ is the Alfv\'en velocity. 
Assuming polarization vectors accurately track the local magnetic field 
direction and transverse velocity perturbations in
the two directions perpendicular to $\hat B$ are comparable, 
the Chandrasekhar-Fermi 
method is rewritten in terms of observational measures,
\begin{equation}
 \sigma_{pol}=f (4\pi\rho_\circ)^{1/2} \sigma_v / B_p 
\end{equation}
where $\sigma_{pol}$ is the dispersion of polarization angles measured 
in radians, $\sigma_v$ is the line of sight velocity dispersion, 
$\rho_\circ$ is the mean density of the gas, and 
the factor, f, accounts for density inhomogeneity and line of sight 
integration.  Ostriker, Stone, \& Gammie (2001) and Padoan etal (2001) 
determine $f\approx$0.4-0.5 from computational simulations. 
The dispersion of measured optical polarization angles within the 
target field is 0.17 radians.  The line of sight velocity dispersion 
determined from the \coa\ data is 0.38 
\kms.  Assuming a mean density of 250 \cc\ and f=0.5, the derived 
mean field strength is 14 ${\mu}G$. 
Thus, our PCA-based estimate of the magnetic field strength in Taurus also 
compares favorably to the value derived by the Chandrasekhar-Fermi method. 

\subsection{The Magnetic Support of the Taurus Cloud Envelope}
The degree to 
which the magnetic field can support a volume against self-gravitational 
collapse is parameterized by the mass to flux ratio with 
respect to the critical value,
$(M/\Phi)_{crit} = 0.16/G^{1/2} $
(Nakano \& Nakamura 1978). 
The magnetic critical index, $\mu$, is the ratio of the 
mass to flux ratio of a volume to this critical value,
\begin{equation}
 \mu = (M/\Phi)/(M/\Phi)_{crit} = 7.6\times10^{-21} N(H_2)/B 
\end{equation}
where $N(H_2)$ is the gas column density in cm$^{-2}$ along field 
lines and B is the magnetic field strength expressed 
in $\mu$G. 
Owing to projections of the magnetic field and the mass distribution 
along 
field lines, the observed index, $\mu_{obs}$, overestimates the true magnetic index.
Assuming random orientations of the magnetic field and flattened gas 
distribution with respect to the observer, one can derive a 
statistical correction to the observed value, $<\mu> = \mu_{obs}/3$ 
to assess whether a volume is super-critical ($<\mu>$ greater than 1) or 
sub-critical ($<\mu>$ less than 1) 
(Heiles \& Crutcher 2005).  

Goldsmith etal (2008) derive the distribution of molecular hydrogen 
over 100 deg$^2$ of the Taurus Molecular Cloud using the \co\ and \coa\ 
J=1-0 data of Narayanan etal (2008).  From the Goldsmith etal (2008) image, 
the mean column density within the Taurus subfield analyzed in this study 
is 1.5$\times$10$^{21}$ cm$^{-2}$. For a magnetic field with 
strength 14 $\mu$G, this column density corresponds to an 
observed magnetic index of 
$\mu_{obs} = 0.81$.  Applying the statistical correction for 
projections, $<\mu> = 0.27$.  This low column density subfield 
within the Taurus cloud is magnetically sub-critical indicative of 
a magnetically supported cloud envelope.  
Such sub-critical, low column density  
envelopes are expected given the exposure 
to the ambient UV radiation field that maintains a sufficient degree of 
ionization to couple the neutral material to ions.  
The ambipolar diffusion time scale is long  
with respect to the dynamical time of the envelope.  While 
star formation within the high density cores and filaments of the
Taurus cloud attest to 
the gravitational collapse and lack of magnetic support within 
localized regions, these occupy a 
small fraction of the mass and area of the cloud.  Goldsmith etal 
(2008) report that 50\% of the mass and 75\% of the area 
of Taurus have molecular 
column densities less than 2$\times$10$^{21}$ cm$^{-2}$.
If this column density regime is similar to the subfield analyzed 
in this study, then 
the Taurus molecular cloud envelope remains magnetically supported.

\section{Summary}
We have developed an analysis method to assess velocity anisotropy within 
interstellar molecular clouds from spectroscopic imaging observations. 
Such anisotropy is predicted from 
theory
of strong MHD turbulence
(GS95).  
The utility of 
our 
method is demonstrated 
using 
MHD simulations with 
varying magnetic field strengths. 
Velocity anisotropy is recovered 
in models with strong magnetic fields
($\beta=0.01$) 
oriented perpendicular to the line-of-sight.
No anisotropy is measured in simulations with the magnetic field
pressure 
more 
comparable to 
the local thermal pressure,
or a few times larger
($\beta =1.0, 0.1$).
The analysis is applied to \co\ J=1-0 emission from a low density
sub-region within
the Taurus molecular cloud.  We detect velocity anisotropy that is aligned 
within $\sim$10 degrees of the local magnetic field direction.  This 
coincidence of the field direction with measured 
anisotropy in small-scale velocity variations
demonstrates a 
strong coupling of the interstellar field with the neutral gas  that may 
result from MHD turbulent flows.  Our estimate of the plane-of-sky
magnetic field strength based on our velocity anisotropy 
analysis is in agreement with the value derived using the
Chandrasekhar-Fermi method.  
Based on our estimated magnetic field strength combined with column
density measurements, we find that the low-density envelope of Taurus,
which comprises 
the bulk of the cloud's mass, is magnetically subcritical.

\acknowledgments
This work was supported by NSF grant AST 0540852  to the Five College
Radio Astronomy Observatory.
ECO is supported by NSF grant AST0507315.

\clearpage
\begin{table}[ht]
\label{table1}
\begin{center}
\caption{MHD Simulation Parameters}
\vspace{7mm}
\begin{tabular}
{ccll}
\hline
Model & $\beta$ & $t/t_s$ & $M_s$ \\
\hline
B2 & 0.01 & 0.07 & 7.4 \\
C2 & 0.10 & 0.04 & 7.6 \\
D2 & 1.00 & 0.05 & 7.2 \\
B3 & 0.01 & 0.19 & 4.9 \\
C3 & 0.10 & 0.09 & 4.9 \\
D3 & 1.00 & 0.09 & 4.9 \\
\hline
\end{tabular}
\end{center}
\end{table}

\newpage
\begin{table}[ht]
\label{table2}
\begin{center}
\caption{Structure Function Parameters: $(S_2(\tau))^{1/2} = v_{\circ} \tau^{\gamma}$  }
\vspace{7mm}
\begin{tabular}
{ccccc}
\hline
Model & $\gamma_\parallel$ & $\gamma_\perp$ & $v_{\circ,\parallel}$ & $v_{\circ,\perp}$ \\

\hline
\hline
 B2 & 0.68 & 0.35 & 0.47 & 1.76 \\
 C2 & 0.53 & 0.49 & 0.80 & 0.89 \\
 D2 & 0.63 & 0.61 & 0.60 & 0.63 \\
 B3 & 0.82 & 0.29 & 0.14 & 1.16 \\
 C3 & 0.49 & 0.44 & 0.74 & 0.92 \\
 D3 & 0.56 & 0.52 & 0.54 & 0.57 \\
\hline
\end{tabular}
\end{center}
\end{table}

\newpage
\begin{table}[ht]
\label{table3}
\begin{center}
\caption{Structure Function Parameters Derived from Model Data Cubes}
\vspace{7mm}
\begin{tabular}
{c|cccc|cccc}
\hline
 & \multicolumn{4}{c}{\co} & \multicolumn{4}{c}{\coa} \\
Model & $\gamma_\parallel$ & $\gamma_\perp$ & $v_{\circ,\parallel}$ & $v_{\circ,\perp}$ & $\gamma_\parallel$ & $\gamma_\perp$ & $v_{\circ,\parallel}$ & $v_{\circ,\perp}$ \\
\hline
\hline
B2 & 0.46 & 0.26 & 0.11 & 0.23 & 0.55 & 0.32 & 0.08 & 0.15 \\
C2 & 0.67 & 0.53 & 0.09 & 0.12 & 0.57 & 0.44 & 0.10 & 0.12 \\
D2 & 0.62 & 0.53 & 0.10 & 0.11 & 0.53 & 0.59 & 0.10 & 0.10 \\
B3 & 0.61 & 0.15 & 0.04 & 0.17 & 0.62 & 0.23 & 0.04 & 0.13 \\
C3 & 0.38 & 0.39 & 0.12 & 0.12 & 0.34 & 0.53 & 0.10 & 0.08 \\ 
D3 & 0.49 & 0.31 & 0.09 & 0.11 & 0.61 & 0.41 & 0.07 & 0.09 \\
\hline
\end{tabular}
\end{center}
\end{table}

\newpage

\newpage

\newpage

\newpage

\newpage

\end{document}